\documentstyle [12pt,twoside]{article}

\newcommand{\bq}{\begin{equation}}
\newcommand{\ba}{\begin{eqnarray}}
\newcommand{\eq}{\end{equation}}
\newcommand{\ea}{\end{eqnarray}}

\def\bo{{\raise.15ex\hbox{\large$\Box$}}}
\def\bob{{\lower.2ex\hbox{\large$\Box$}}}

\def\TH{{\raise.2ex\hbox{$\displaystyle \bigodot$}\mskip-4.7mu \llap H \;}}

\catcode`@=11
\def\underline#1{\relax\ifmmode\@@underline#1\else
        $\@@underline{\hbox{#1}}$\relax\fi}
\catcode`@=12
 
\begin{document}

\input{psfig}
\hfill LA-UR-95-3393, LBL-37786
\centerline{\large{\bf Dissipation and Decoherence in Mean Field
Theory}}

\vspace{1.5cm}
 	 
\centerline{\bf Salman Habib$^1$, Yuval Kluger$^2$, Emil Mottola$^1$,
and Juan Pablo Paz$^3$}  

\vspace{1cm} 

\centerline{\em $^1$Theoretical Division} 
\centerline{\em Los Alamos National Laboratory}
\centerline{\em Los Alamos, NM 87545, USA}

\vspace{.5cm} 

\centerline{\em $^2$Nuclear Science Division, MS 70A-3307}
\centerline{\em Lawrence Berkeley National Laboratory}
\centerline{\em Berkeley, CA 94720, USA} 

\vspace{.5cm}
 
\centerline{\em $^3$Departamento de Fisica, FCEN, UBA}
\centerline{\em Pabellon 1, Ciudad Universitaria}
\centerline{\em 1428 Buenos Aires, Argentina}

\vspace{1.5cm}

\centerline{\bf Abstract}

The time evolution of a closed system of mean fields and
fluctuations is Hamiltonian, with the canonical variables
parameterizing the general time-dependent Gaussian density matrix of
the system. Yet, the evolution manifests both quantum decoherence and
apparent irreversibility of energy flow from the coherent mean fields
to fluctuating quantum modes. Using scalar QED as an example we show
how this collisionless damping and decoherence may be understood as
the result of {\em dephasing} of the rapidly varying fluctuations and
particle production in the time varying mean field.
\newpage

Mean field methods have a long history in such diverse areas as atomic
physics (Born-Oppenheimer), nuclear physics (Hartree-Fock), condensed
matter (BCS) and statistical physics (Landau-Ginzburg), quantum optics
(coherent/squeezed states), and semiclassical gravity.  Because no
higher than second moments of the fluctuations are incorporated, the
mean field approximation is related to a Gaussian variational ansatz
for the wave function of the system. The broad applicability of the
approximation, as well as the variety of different approaches to it in
the literature makes it worthwhile to exhibit its general features
unobscured by the particulars of specific applications. Accordingly,
our first purpose in this Letter is to demonstrate the equivalence of
the time-dependent mean field approximation to the general Gaussian
ansatz for the mixed state density matrix $\rho$, and to underline its
Hamiltonian structure.

The Hamiltonian nature of the evolution makes it clear from the outset
that the mean field approximation does {\em not} introduce dissipation
or time irreversibility at a fundamental level. Nevertheless, typical
evolutions seemingly manifest an arrow of time, in the sense that
energy flows from the mean field to the fluctuations without returning
over times of physical interest [Fig. 1]. Closely connected to this
{\em effective} dissipation is the phenomenon of quantum decoherence
\cite{deco}, {\em i.e.}, the suppression with time of the overlap
between wave functions corresponding to two different mean field
evolutions [Fig. 2]. This decoherence is the reason why quantum
superpositions of different mean field states are difficult to observe
in nature, and is crucial to understanding the quantum to classical
transition in macroscopic systems.  Our second aim in this Letter is
to present an explicit example of a quantum field theory (scalar QED)
treated in mean field approximation where these effects are observed,
and to provide a clear physical explanation of the behavior in terms
of dephasing of the fluctuations, {\em i.e.}, the averaging to zero of
their rapidly varying phases on time scales short compared to the
collective motion of the mean field(s).

To expose the general structure of the time-dependent mean field
(or Gaussian) approximation consider first a
one-dimensional harmonic oscillator with Hamiltonian, 
\begin{equation}
H_{osc} (q, p; t) = {1 \over 2}\left(p^2 + \omega^2(t)\, q^2\right)
\label{Ham}
\end{equation}
where the frequency $\omega(t)$ is a smooth function of time,
otherwise unspecified for the moment. The most general Gaussian ansatz
for the mixed state normalized density matrix may be presented as
\begin{eqnarray}
\langle x'|\rho|x\rangle &=& (2\pi \xi^2)^{-{1\over 2}}\exp \biggl\{
i{\bar p\over\hbar} (x'-x)-{\mu^2 + 1\over 8 \xi^2}\left[ (x'-\bar
q)^2 + (x-\bar q)^2\right] \nonumber \\  
&& + i\,{\eta \over 2 \hbar\xi} \left[ (x'-\bar q)^2 - (x- \bar
q)^2\right] + {\mu^2 - 1\over 4 \xi^2} (x'- \bar q)(x- \bar q)
\biggr\}~,\label{gauss} 
\end{eqnarray}
in the coordinate representation. The five parameters ($\bar q, \bar
p, \xi, \eta, \mu $) of this Gaussian may be identified with the two
mean values, $\bar q = \langle q \rangle \equiv {\rm Tr}(q \rho),~\bar
p = \langle p \rangle \equiv {\rm Tr}(p \rho)$, and the three
symmetrized variances via
\begin{eqnarray}
\langle(q-\bar q)^2 \rangle = \xi^2~,~~&&~\langle(pq + qp - 2\bar q
\bar p)\rangle =  2\xi \eta~,\nonumber \\  
\langle(p-\bar p)^2 \rangle\ &=& \eta^2 + {\hbar^2 \mu^2 \over 4
\xi^2}\ . 
\label{var}
\end{eqnarray}
The one anti-symmetrized variance is fixed by the commutation
relation, $[q, p] = i\hbar$. The parameter $\mu$ measures the degree
to which the state is mixed: ${\rm Tr}~\rho^2 =\mu^{-1} \le 1$, the
equality holding for pure states. If the state is pure, $\rho =
|\psi\rangle\langle\psi|$, and only two of the three symmetrized
variances in (\ref{var}) are independent.

The Gaussian ansatz for the density matrix is preserved under time
evolution. In the Schr\"odinger picture $\rho$ evolves according to
the Liouville equation, $\dot \rho = -i[H,\rho]$. Substitution of the
Gaussian form (\ref{gauss}) into this equation with Hamiltonian
(\ref{Ham}) and equating coefficients of $x$, $x'$, $x^2$, $x'^2$ and
$xx'$ gives five evolution equations for the five parameters
specifying the Gaussian,
\begin{equation}
\begin{array}{lc}
\dot{\bar q} = \bar p \ ;\qquad & \dot{\bar p} = - \omega^2 (t) \bar q
\\    
\dot{\xi} = \eta \ ; \qquad & \dot{\eta} = - \omega^2 (t) \xi +
{\hbar^2 \mu^2 \over 4 \xi^3}  
\end{array}
\label{evol}
\end{equation}
and $\dot{\mu} = 0$. Since $\mu$ is a constant and the von Neumann
entropy, $-{\rm Tr}~\rho\ln \rho$ of the state (\ref{gauss}) is a
(monotonic) function of $\mu$ alone, this quantity is also a constant
of the motion.  Evolution equations for the diverse applications of
the time-dependent mean field approximation reduce to (multiple copies
of) equations of precisely the general form of (\ref{evol}), with
$\omega^2(t)$ a different self-consistently determined function of the
coordinates and time, depending on the application. This establishes
the equivalence between mean field methods and Gaussian density
matrices for all evolutions of the form of Eqns. (\ref{evol}). We give
explicit examples below.

An essential property of the evolution equations (\ref{evol}) is that
they are Hamilton's equations (hence, time reversible) for an
effective classical Hamiltonian \cite{classham}, with $\eta$ playing
the role of the momentum conjugate to $\xi$,
\begin{equation}
H_{eff}(\bar q, \bar p ; \xi , \eta) = {\rm Tr}(\rho H) 
= {1\over 2}\left(\bar p^2 + \eta^2\right) + V_{eff}\ , 
\label{effH}
\end{equation}
and $V_{eff}(\bar q, \xi)$ depending on the particular form of
$\omega^2(\bar q(t),\xi (t);t)$. For example, if the original system
is an anharmonic double well with quantum Hamiltonian
\begin{equation}
H(q, p) = {1 \over 2}p^2 +  {\lambda \over 4}\left(q^2 -
v^2\right)^2~,  
\label{anhH}
\end{equation}
then the (large N) mean field equations of motion are identical to
Eqs. (\ref{evol}) with $\omega^2(t) = \lambda (\bar q^2 (t) + \xi^2(t)
- v^2)$. In this case, 
\begin{equation}
V_{eff}(\bar q, \xi) =  {\lambda\over 4}\left(\bar
q^2 + \xi^2 - v^2\right)^2 + {\hbar^2 \mu^2 \over 8 \xi^2}~.
\label{effV}
\end{equation}
The resulting evolution equations (\ref{evol}) are now quite
non-linear (and chaotic). The last ``centrifugal'' term in the
effective potential is a manifestation of the quantum uncertainty
principle which prevents the Gaussian width $\xi$ from shrinking to
zero.
 
The unitary operator $U(t)$ which
effects the time evolution of the density matrix (\ref{gauss}),    
\begin{equation}
\rho (t) = U(t) \rho (0) U^{\dagger}(t)\, , \quad U(t) = \exp \left(
-i\int_0^t H dt \right)
\label{unitev} 
\end{equation}
is given explicitly in the coordinate basis by
\begin{equation}
\langle x'|U(t)|x\rangle = (2\pi i\hbar v(t))^{-{1\over
2}}\exp \left\{ {i\over 2\hbar v(t)}\left(u(t)x^2 + \dot v(t)x'^2 -
2xx'\right)\right\}  \label{unit}
\end{equation}
in terms of the two linearly independent solutions to the classical
evolution equation, 
\begin{equation}
\left({d^2 \over dt^2} + \omega^2 (t)\right) \left( \begin{array}{c}
u\\v 
\end{array}\right) = 0\ ; \qquad \begin{array} {l} u(0) = \dot v (0) = 
1\\ \dot u(0) = v(0) = 0~. 
\end{array}
\label{evolop}
\end{equation} 

The Gaussian dynamics may be expressed as well by means of a Fock
representation of the time dependent Heisenberg operators, 
\begin{eqnarray}
q(t) &=& U^{\dagger}(t)\,q(0)\,U(t) =\bar q(t) + a f(t) + a^{\dagger} 
f^*(t) \nonumber\\ 
p(t) &=& U^{\dagger}(t)\,p(0)\,U(t) =\bar p(t) + a \dot f(t) +
a^{\dagger} \dot f^*(t)
\label{abas} 
\end{eqnarray}
where $[a,a^{\dagger}] =1$. The complex mode functions $f$ satisfy the
evolution equation (\ref{evolop}) and the Wronskian condition,
\begin{equation}
f \dot f^* - \dot f f^*  = i \hbar
\label{Wron}
\end{equation}
showing that Gaussian time evolution is essentially classical, with
$\hbar$ appearing only in the time independent condition (\ref{Wron})
enforcing the quantum uncertainty relation.  There is considerable
latitude to redefine $f$ by the Bogoliubov transformation $f
\rightarrow \cosh\gamma\ e^{i(\theta +\phi)} f + \sinh \gamma\
e^{i(\theta -\phi)} f^*$ without affecting the Wronskian condition
(\ref{Wron}). Such Bogoliubov transformations form a noncompact Lie
group, the metaplectic group $Mp(2)$, which is a double covering of
the symplectic group of classical Hamiltonian dynamics, $Sp(2) \cong
SU(1,1)$, whose Lie algebra is generated by the three symmetric
bilinears $aa$, $a^{\dagger}a^{\dagger}$, and $aa^{\dagger} +
a^{\dagger}a$. If the $a$ and $a^{\dagger}$ operators are appended to
these three, the algebra again closes upon itself, forming a five
parameter group $IMp(2)$ \cite{rgl}. The unitary evolution
(\ref{unitev}), (\ref{unit}) of the Gaussian density matrix
(\ref{gauss}) is an explicit representation of this group.
 
The group structure can be exploited to choose a basis in which all
expectation values vanish, except  
\begin{equation}
\langle a^{\dagger}a\rangle=\langle aa^{\dagger}\rangle - 1 \equiv
N \ge 0~. 
\label{bogl}
\end{equation}
The Gaussian density matrix is diagonal in the corresponding
$a^{\dagger}a$ time-independent number basis, 
\begin{equation}
\langle n'|\rho|n\rangle = {2\delta_{n'n}\over \mu +1}\left({\mu
-1\over \mu +1}\right)^n \ ,    
\label{diag}  
\end{equation}
with $\mu = 2 N + 1$ and $\xi^2(t) = \mu \vert f(t)\vert^2$. Upon
identifying $\mu = \coth (\hbar \omega/2kT)$, the diagonal form
(\ref{diag}) will be recognized as a thermal density matrix at
temperature $T$.  The pure state Gaussian wave function ($\mu =1$)
corresponds therefore to a coherent, squeezed zero temperature vacuum
state. The smoothness of the finite temperature classical limit
$\hbar\mu \rightarrow 2kT/\omega$ as $\hbar\rightarrow 0$, $\mu
\rightarrow \infty$ shows that quantum and thermal fluctuations are
treated by the mean field approximation in a unified way.
  
By making another $IMp(2)$ group transformation it is always possible
to diagonalize (\ref{Ham}) at any given time, bringing the quadratic
Hamiltonian into the standard harmonic oscillator form, $H_{osc} =
{\hbar\omega\over 2}\left(\tilde a \tilde a^{\dagger} + \tilde
a^{\dagger}\tilde a\right)$ with $\tilde a$ time dependent. This time
dependent basis is defined by the relations,
\begin{eqnarray}
q(t) &=& \tilde a \tilde f + \tilde a^{\dagger} \tilde f^*
\qquad p(t) = -i\omega \tilde a \tilde f + i\omega \tilde a^{\dagger}
\tilde f^* \nonumber \\ 
\tilde f (t) &=& \sqrt {\hbar\over 2 \omega (t)}\, \exp \left(
-i\int_0^t \, dt' \omega(t')\,\right)~,  
\end{eqnarray}
in place of (\ref{abas}).  In the $\tilde a^{\dagger} \tilde a$ number
basis, $\rho$ is no longer diagonal, $\langle\tilde a \rangle$,
$\langle \tilde a\tilde a\rangle$, {\em etc.}, are non-vanishing, and
$\tilde N \equiv\langle\tilde a^{\dagger}\tilde a\rangle \neq N$ in
general, becoming equal only in the static case of constant $\omega$.

If $\omega(t)$ varies slowly in time, an adiabatic invariant may be
constructed from the Hamilton-Jacobi equation corresponding to the
effective classical Hamiltonian (\ref{effH}).  By a simple quadrature
we find the adiabatic invariant,
\begin{equation}
{W\over 2\pi\hbar} = {\langle H\rangle\over \hbar\omega} - {\mu\over 2} 
=\tilde N - N\ . 
\end{equation}
Since $N$ is time independent, $\tilde N (t)$ is an adiabatic
invariant of the evolution. On the other hand, the phase angle
conjugate to the action variable $W$ varies rapidly in time.  Since
the diagonal matrix elements of $\rho$ in the $\tilde N$ basis are
independent of this phase angle, they are slowly varying, whereas the
{\it off-diagonal} matrix elements of $\rho$ in this basis (which
depend on the phase angle) are {\it rapidly} varying functions of
time. If we are interested only in the effects of the fluctuations on
the more slowly varying mean fields it is natural to define an {\it
effective} density matrix $\rho_{eff}(t)$ by {\it time-averaging} the
density matrix (\ref{gauss}), thereby truncating $\rho$ to its
diagonal elements only, in the adiabatic $\tilde N$ basis
\cite{Kand}. Clearly, for this truncation to be justified there must
be very efficient phase cancellation, {\it i.e. dephasing}, either by
averaging the fluctuations over time or by summing over many
independent fluctuating degrees of freedom at a fixed time.

Obtaining the general form of the diagonal matrix elements of $\rho$
in the $\tilde N$ basis is straightforward but the result is rather
unwieldy. Here we restrict our attention to the case of a pure state
with vanishing $\bar q$ mean field. Using the methods of
Ref. \cite{Brown}, one finds simply
\begin{equation}
\langle \tilde n=2\ell|\rho|\tilde n = 2\ell\rangle
\bigg\vert_{_{\stackrel{\mu =1}{\bar q =\bar p = 0}}} 
= {(2\ell -1)!!\over 2^{\ell}\, \ell !} {\rm sech} \gamma
\,\tanh^{2\ell} \gamma 
\label{diagrho} 
\end{equation}
with $\rho_{\tilde n\tilde n} = 0$ for $\tilde n$ odd and $\gamma (t)$
the parameter of the Bogoliubov transformation between the $a$ and
$\tilde a$ bases. It is given explicitly by $\sinh^2 \gamma = \tilde N
= \vert \dot f + i \omega f\vert^2 /2 \hbar\omega $.

Decoherence is also addressable within the same framework. Consider
the case where $\omega(t)$ is a function of one external mean field
degree of freedom $A(t)$. If only the evolution of $A$ is of interest,
then the fluctuating modes described by $f(t)$ may be treated as the
``environment.'' To solve for the evolution of the reduced density
matrix of $A$, one needs to compute the influence functional. This is
a functional of two trajectories $A_1(t)$ and $A_2(t)$ (corresponding
to two different evolution operators $U_1(t)$ and $U_2(t)$), and is
defined by
\begin{equation}
F_{12}(t) \equiv \exp(i\Gamma_{12}(t))\equiv {\rm
Tr}\left(U_1(t)\rho(0)U_2^{\dagger}(t)\right). 
\label{if}
\end{equation}
Explicit evaluation may be carried out using (\ref{unit}). Restricting
again to pure states with vanishing $\bar q$ mean fields we find
\begin{equation}
\Gamma_{12}\bigg\vert_{_{\stackrel{\mu =1}{\bar q =\bar p = 0}}}
=-{i\over 2}\ln \left\{{i\hbar\over|f_1f_2|}\left({f_1f_2^*\over 
f_1\dot{f}_2^*-\dot{f}_1f_2^*}\right)\right\} 
\label{gamexp}
\end{equation}
in terms of the two sets of mode functions $f_1(t)$ and $f_2(t)$ which
satisfy (\ref{evolop}) and (\ref{Wron}).  This $\Gamma_{12}$ is
precisely the closed time path (CTP) effective action functional which
generates the connected real time $n$-point vertices in the quantum
theory \cite{largeN}.  For a pure initial state, the absolute value of
$F_{12}$ measures the overlap of the two different evolutions at some
time $t$, beginning with the same initial $\vert\psi(0)\rangle$.  In
mean field theory, instead of evaluating $\Gamma_{12}$ for two
arbitrary trajectories, the evaluation is over trajectories determined
by the {\it self-consistent} evolution of the closed system, beginning
with two different initial mean fields.

For an explicit field theoretic example consider scalar QED with no
scalar self-coupling. In the large $N$ limit, the evolution of
electric fields and charged matter field fluctuations may be described
in the self-consistent mean field or Gaussian approximation
\cite{largeN}. For a spatially homogeneous electric field in the $z$
direction, in the gauge $\vec A = A(t) \hat z$, the time evolution
equations read
\begin{eqnarray}
\ddot A(t) = \langle j_z(t)\rangle = {2e\over V}\sum_{\vec k}
\left(k_z - eA(t)\right)&& |f_{\vec k}(t)|^2 (2N(\vec k) + 1)\nonumber\\ 
\left[ {d^2\over dt^2} + \left(\vec k - e\vec A(t)\right)^2 + m^2
\right] f_{\vec k}(t) &&= 0 
\label{sqed}
\end{eqnarray}
with the Wronskian condition (\ref{Wron}) holding for every discrete
wave number $\vec k$ in the finite volume $V$. In field theory there
are an infinite number of quantum fluctuating modes $f_{\vec k}$ of
the charged scalar field, each varying rapidly in time with its own
characteristic frequency, $\omega^2_{\vec k}(t) = (\vec k - e\vec
A(t))^2 + m^2$.  The Gaussian density matrix is an infinite product of
Gaussians each of the form (\ref{gauss}) and there are an infinite
number of Bogoliubov parameters $\gamma (\vec k ; t)$, one for each
$\vec k$.  The equations of motion (\ref{sqed}) are again Hamiltonian
in structure.  In this case the effective classical Hamiltonian,
\begin{equation}
H_{eff} = {E\over 2}^{\!2} + {1\over V}\sum_{\vec k} \left[\vert \dot
f_{\vec k}\vert^2 + \omega^2_{\vec k}\,\vert f_{\vec k}\vert^2
\right](2N(\vec k) + 1)
\end{equation} 
describes charged particle production in the electric field $E = -\dot
A$ by the Schwinger mechanism and the effects of the current $\langle
j_z(t)\rangle $ generated by these charged particles back on the
electric field, through the semiclassical Maxwell equation in
(\ref{sqed}).  The mean value of the scalar field itself is zero so
that we may use the expressions (\ref{diagrho}) and (\ref{gamexp}) for
the effective density matrix and decoherence functional of the charged
field fluctuations. The values, $\rho_{2\ell}$ given by
(\ref{diagrho}) are then the probabilities of observing $\ell$ charged
particle pairs in the adiabatic $\tilde N$ basis. The diagonal matrix
elements of $\rho$ for odd $\tilde n$ vanish because particles can
only be created in pairs from the vacuum.

\begin{figure}
\centerline{\psfig{figure=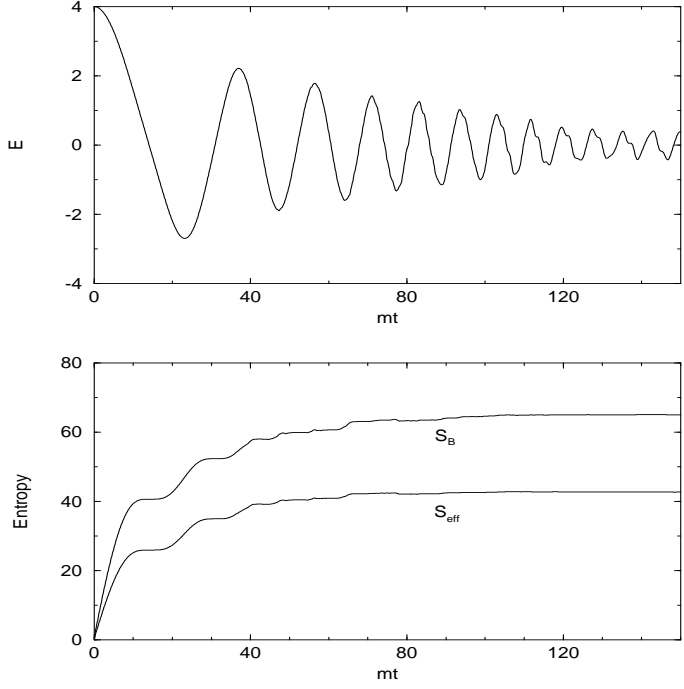,height=9.0cm,width=9.0cm}}
\caption[Figure 1]{\small{Evolution of the electric field, and
the Boltzmann and effective entropies. The electric field is 
expressed in units of $E_c= m^2 c^3/e\hbar$. Pair creation is
rapid when $|E| > E_c$.}}
\end{figure}

\begin{figure}
\centerline{\psfig{figure=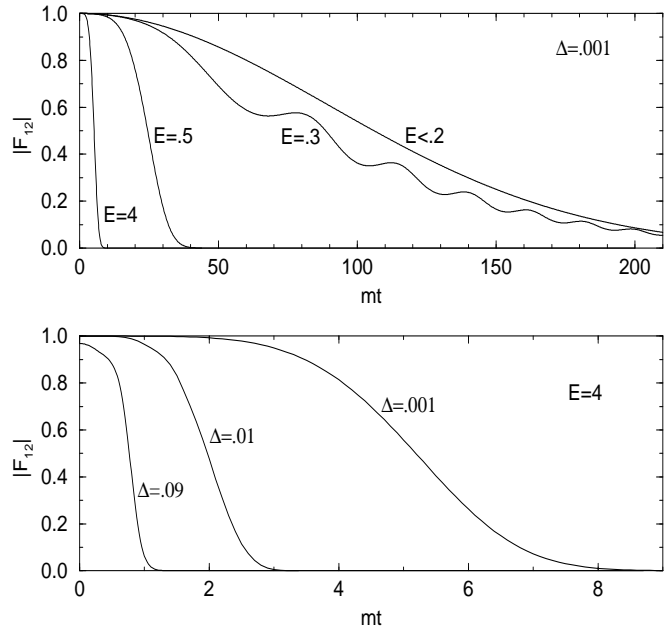,height=9.0cm,width=9.0cm}}
\caption[Figure 1]{\small{Absolute value of the decoherence functional
as a function of time. The two field values are $E$ and
$E-\Delta$. The top figure shows (for fixed $\Delta$) the sharp
dependence of decoherence on particle production when $|E| \ge .2\, E_c$. 
The second illustrates the relatively milder dependence on $\Delta$.}}  
\end{figure}

In this specific model we present numerical results (in $1+1$
dimensions with vacuum initial conditions) on damping and decoherence
of the mean electric field in Figs. 1 and 2. The plasma oscillations
of the mean electric field on timescales long compared to
$\omega_{\vec k}^{-1}$ are clearly seen, as well as the apparently
irreversible flow of energy from the electric field towards the
charged particle modes. In Fig. 1 we plot both the Boltzmann entropy
and the von Neumann entropy of the effective density matrix, for
comparison. These are defined by
\begin{eqnarray}
S_{B} &\equiv &\sum_{\vec k} \left\{ \left( \tilde N(\vec k) 
+ 1 \right) \ln \left( \tilde N(\vec k) + 1\right) - \tilde N(\vec
k) \ln \tilde N(\vec k) \right\}\nonumber\\ 
S_{eff} &\equiv& - {\rm Tr}\rho_{eff}\ln\,\rho_{eff} = -\sum_{\vec k}
\sum_{\ell = 0}^{\infty} \rho_{2\ell}(\vec k) \ln\, \rho_{2\ell} (\vec
k)   
\end{eqnarray}
respectively. Both display general increase during intervals of
particle creation \cite{CalHu}, when the electric field is
sufficiently strong for the Schwinger pair creation mechanism to be
effective. Neither quantity is a strictly monotonic function of time
(no $H$-theorem). Since the charged particle modes $f_{\vec k}$
interact with the mean electric field but not directly with each
other, the effective damping observed is certainly {\it
collisionless}, and may be understood as due to dephasing similar to
that responsible for Landau damping of collective modes in classical
electromagnetic plasmas. The entropy $S_{eff}$ of the effective
density matrix provides a precise measure of the information lost by
treating the phases as random. The Boltzmann ``entropy" would be
expected to equal $S_{eff}$ only in true thermodynamic equilibrium,
which is not achieved in the collisionless approximation of
Eqns. (\ref{sqed}). Otherwise we see from Fig. 1 that $S_B$ generally
overestimates the amount of information lost by phase averaging.

That decoherence is closely related to the same dephasing of the
particle modes is seen most clearly by comparing the absolute value of
$F_{12}$ for different initial electric fields. Decoherence is very
slow for electric fields less than the Schwinger pair production
threshold but becomes very rapid above it \cite{Kief}. This shows the
strong dependence of the decoherence process on the same particle
production by the mean field.

The authors wish to acknowledge several helpful discussions 
with F. Cooper and A. Kovner.

\end{document}